\documentclass[twocolumn,showpacs,preprintnumbers,superscriptaddress,amsmath,floatfix,amssymb,prd]{revtex4}
\usepackage{graphicx}
\usepackage[colorlinks=true]{hyperref}
\usepackage{amsfonts,amsmath,amsxtra}

\newcommand{\beq}[1]{\begin{eqnarray}\label{#1}}
\newcommand\eeq {\end{eqnarray}}
\newcommand\bqa {\begin{eqnarray}}
\newcommand\eqa {\end{eqnarray}}

\newcommand{\bear}{\begin{array}}
\newcommand{\enar}{\end{array}}

\newcommand{\R}{\mathbb{R}}
\newcommand{\C}{\mathbb{C}}
\newcommand{\K}{\mathcal{K}}

\newcommand{\Z}{\mathbb{Z}}

\newcommand{\N}{\mathbb{N}}

\begin{document}
\def\le{\langle}
\def\re{\rangle}
\def\dg{^{\dag}}
\def\K{{\cal K}}
\def\n{{\cal N}}
\font\maj=cmcsc10 \font\itdix=cmti10
\def\N{\mbox{I\hspace{-.15em}N}}
\def\H{\mbox{I\hspace{-.15em}H\hspace{-.15em}I}}
\def\1{\mbox{I\hspace{-.15em}1}}
\def\pN{\rm{I\hspace{-.1em}N}}
\def\Z{\mbox{Z\hspace{-.3em}Z}}
\def\pZ{{\rm Z\hspace{-.2em}Z}}
\def\R{{\rm I\hspace{-.15em}R}}
\def\pR{{\rm I\hspace{-.10em}R}}
\def\C{\hspace{3pt}{\rm l\hspace{-.47em}C}}
\def\Q{\mbox{l\hspace{-.47em}Q}}
\def\b{\begin{equation}}
\def\e{\end{equation}}
\def\bee{\begin{enumerate}}
\def\wt{\widetilde}

\title{Non-Linear Trans-Planckian Corrections of Spectra\\ due to the Non-trivial Initial States}
\author{E. Yusofi}
\email{e.yusofi@iauamol.ac.ir}
\affiliation{Department of Physics,
Science and Research Branch, Islamic Azad University, Tehran, Iran}
\author{M. Mohsenzadeh}
\email{mohsenzadeh@qom-iau.ac.ir}
\affiliation{Department of Physics, Qom Branch, Islamic Azad University, Qom, Iran}
\date{\today}
\begin{abstract}
Recent Planck results motivated us to use non-Bunch-Davies vacuum. In this paper, we use the excited-de Sitter mode as non-linear initial states during inflation to calculate the corrected spectra of the initial fluctuations of the scalar field. First, we consider the field in de Sitter space-time as background field and for the non-Bunch-Davies mode, we use the perturbation theory to the second order approximation. Also, unlike conventional renormalization method, we offer de Sitter space-time as the background instead Minkowski space-time. This approach preserve the symmetry of curved space-time and stimulate us to use excited mode. By taking into account this alternative mode and the effects of trans-Planckian physics, we calculate the power spectrum in standard approach and Danielsson argument. The calculated power spectrum with this method is finite, corrections of it is non-linear, and in de Sitter limit corrections reduce to linear form that obtained from several previous conventional methods.
\end{abstract}
\maketitle

\section{Introduction}
As a causal manner, inflation scenario can be explain density
perturbations originating in areas outside the horizon in the very
early universe\cite{Lin1, Bau2}. These perturbations actually planted the
seeds of the current observable structure in the universe. During
inflation, the physical size of any perturbation mode has grown
faster than the horizon, cross of  it and freezes. After the end of
inflation, these super-horizon perturbed mode re-enter into the
horizon and along with the expansion, form into galaxies and
clusters under the gravitational force\cite{Muk3, Muk4}. Naturally,
this causal relations in the sub-horizon radius allows us to
delineate their initial amplitude by considering that perturbations start from quantum fluctuations of their vacuum.\\
 The measurable radiation from the Big Bang is cosmic microwave background radiation(CMBR), that
provides a snapshot of the early universe and our main notion of the
early universe has improved significantly over the survey of it.
Most of the inflationary models show an anisotropy in the
temperature in the CMBR map \cite{Spe5} and it could be indicate
the quantum origin of the universe and come from the primordial quantum fluctuations of the fields in the initial vacuum state.
  It may also be related to effects of "trans-Planckian physics" beyond the
  Planck scale \cite{Dan6, Sen7}. \\
Firstly, the trans-Planckian physics effects in inflation was
introduced in \cite{Kis8}. Dispersion relations and their concepts
in various form have been greatly studied in \cite{Mar9, Mar10, Mar11}. Another
way to discuss the trans-Planckian physics emanated from the quantum
gravity and Non-commutative Geometry of space-time, such as string
theory\cite{Kem12, Kem13, Kem14}. In the another notable attempt, Danielsson
focused on the choice of vacuum and he has used $\alpha$-vacuum
state in the inflationary background \cite{Dan6}.
\\
By consideration of a cutoff length $\Lambda^{-1}$, the conventional
predictions of inflation get corrected due to the finite effects of
the expansion of the universe \cite{Dan6}. These modifications can
be expanded as a series in $ H/\Lambda$. The order of the
correction to the power spectrum obtained by Danielsson \cite{Dan6}
was $ H/\Lambda$. In \cite{Kal11} using the method of
effective field theory the authors found that the correction was $
(H/\Lambda)^{2} $, and when the mode are initially created
by adiabatic vacuum state, the authors found that the correction of
power spectrum was $(H/\Lambda)^{3} $\cite{Nie12}. Some
other works has been performed in \cite{Alb13, Alb14, Alb15, Alb16, Alb17, Alb18}. The complete
analysis of these different orders of correction has discussed in \cite{Mar14, Mar15}.\\
Cosmic inflation is described by nearly de Sitter(dS) space-time. In flat space-time, there exists an unique and well-defined vacuum
state, but in curved space-time, the meaning of vacuum is not very
clear and exists ambiguity in the choice of vacuum\cite{Bir16}. If
we consider universe as exact dS space during inflation,
there exists a concrete set of vacuum states invariant under the
symmetry group of the dS space-time. However, as we know, an
inflating universe is not exact dS space-time but it may be
dS space-time in the first approximation \cite{Lin1}. Since, very early universe stood in the high energy area, it may seem more
logical using of higher order perturbations. Therefore, in
this work, we exploit of perturbation theory and consider \emph{non-linear excited- dS mode} instead linear Bunch-Davies(BD) mode as the fundamental mode during inflation \cite{man17, man18}. The main point of this paper is that the order of corrections to the power spectrum changes
if we consider non-linear initial vacuum mode. To achieve this goal, we use the excited version of $\alpha$-vacuum and we generalize Danielsson work in the context of $\alpha$-vacuum\cite{Dan6}.  \\
    The layout of paper is as follows: In Sec. 2., we briefly
    recall the definition of standard power spectrum and we review
the calculation of standard power spectrum in BD vacuum mode
and $\alpha$-vacuum with trans-Planckian effects. Sec. 3. is the main work of this paper. In 3.1. some motivations for
our offered vacuum mode is introduced. In 3.2. and 3.3., we calculate the
corrected power spectrum with excited-dS vacuum and excited- $\alpha-vacuum$. Conclusions
is given in the final section.

\section{Standard Approach for Calculation of Power Spectrum}

The following metric is used to describe the universe during the
inflation: \b \label{equ1}
ds^{2}=dt^{2}-a(t)^{2}{d\textbf{x}}^{2}=a(\eta)^{2}({d\eta}^2-{d\textbf{x}}^2),
\e
where for dS space, the scale factor is given by $a(t)=\exp(Ht)$ and
$a(\eta)=-\frac{1}{H\eta}$. $\eta$ is the conformal time and $H$ is
the Hubble constant. There are some models of inflation but the
popular and simple one is the single field inflation in which a
minimally coupled scalar field (inflaton) is considered in inflating
background:
\b \label{equ2}
S=\frac{1}{2}\int d^4x\sqrt{-g}\Big(R-(\nabla \phi)^2-m^2\phi^2\Big),
\e
where $8{\pi}G = \hbar =1$. The corresponding inflaton field
equation in Fourier space is given by:
\b \label{equ3}
{\phi''}_{k}-\frac{2}{\eta}{\phi'}_{k}+(k^{2}+a^2m^2)\phi_{k}=0,
 \e
where prim is the derivative with respect to conformal time $\eta$.
For the massless case, with the rescaling of $\upsilon_{k}=a\phi_{k},$ equation (\ref{equ3}) becomes
\b
 \label{equ4} {\upsilon''}_{k}+(k^{2}-\frac{\ddot a}{a})\upsilon_{k}=0.
  \e
The general solutions of this equation can be written as
\cite{Kal11, vac19}:
\b \label{gen5}
\upsilon_{k}=A_{k}H_{\mu}^{(1)}(|k\eta|)+B_{k}H_{\mu}^{(2)}(|k\eta|),
 \e
where $ H_{\mu}^{(1, 2)} $ are the Hankel functions of the first and second kind, respectively \cite{Kal11}. For the large values of $|k\eta|$ the above Hankel functions have the expansion to the second order approximation in the following form\cite{vac19, Abr26},
\b \label{gen6}
H_{\mu}(|k\eta|)\approx\sqrt{\frac{2}{\pi{|k\eta|}}}\big[1-{i}\frac{4\nu^2-1}{8{k\eta}}+\bigcirc(\frac{1}{{k\eta}})^{2}\big]
 \times{exp}[-{i}{k\eta}].
 \e
Note that for far past we consider $|k\eta|=-k\eta$.
\subsection{Power Spectrum with Exact dS Vacuum}
Consider the dS limit (H =constant), and
 \b \label{mod6}
\frac{\ddot{a}}{a}=\frac{2}{\eta^{2}},
 \e
In a pure dS background, we therefore wish to solve the mode equation
\b
\label{equ7}  \upsilon''_{k}+(k^{2}-\frac{2}{\eta^{2}})\upsilon_{k}=0.
\e
 and the exact solution of (\ref{equ7}) becomes \cite{Bau2},
 \b \label{mod8}
\upsilon_{k}=\frac{A_{k}}{\sqrt{2k}}(1-\frac{i}{k\eta})e^{-ik\eta}+\frac{B_{k}}{\sqrt{2k}}(1+\frac{i}{k\eta})e^{ik\eta},
\e  where $A_k$ and $B_k$ are Bogoliubov coefficients. In general, this set of
vacua (labelled by $\alpha$) is used and written by $\alpha$-vacuum.
However, the free parameters $ A_{k} $ and $ B_{k} $ can be fixed to
unique values by considering the quantization condition
$i({\upsilon^{*}_{k}}\acute{\upsilon_{k}}-\upsilon^{*}_{k}\acute{}\upsilon_{k})=1$
together with the sub-horizon limit $ {|k\eta|}\gg1 $, and leads to the
unique Bunch-Davies(BD) vacuum \cite{Bun20} by setting $ B_{k}=0 $
and $ A_{k}=1 $ in (\ref{mod8})
\b \label{mod9}
\upsilon_{k}^{BD}=\frac{1}{\sqrt{2k}}(1-\frac{i}{k\eta})e^{-ik\eta}.
\e
For any given mode $\upsilon_k$, the two-point function in Hilbert space
is defined by: \b \label{two10}
\langle{\phi^{2}}\rangle=\frac{1}{(2\pi)^{3}}\int\frac{|\upsilon_{k}|^{2}}{a^{2}}d^{3}{k}.
\e
Then from (\ref{mod9}) and (\ref{two10}) one can write:
 \b \label{two11}
\langle \phi^{2}\rangle= \frac{1}{(2\pi)^{3}}\int {d^{3}}k[\frac{1}{2ka^{2}}+\frac{H^{2}}{2k^{3}}].
\e
 The usual contribution from vacuum fluctuations in Minkowski
space-time, i.e. the first term, is divergent. This infinity can be eliminated after the renormalization
\cite{Lin1}. One of the simple methods to automatic renormalization device in curved space-time, is
discussed in \cite{Tak21, Tak22, Tak23, Tak24, Tak25, Tak26, Tak27}. Then the renormalized power spectrum for
the scalar field fluctuations is calculated as \cite{Lin1, Sta22}:
\b
 \label{pow12} P_{\phi}(k)=(\frac{H}{2\pi})^{2}.
 \e
Since during the inflationary era, background space-time is
considered curved, it is better to be a general solution for the
wave equation (\ref{equ4}), include both positive and negative norm. The
$\alpha$-vacuum is obtained from (\ref{mod8}) in general case, in
which $ B_{k}\neq 0 $ and $ A_{k}\neq0 $. In section (2.3) of
\cite{Dan6}, Danielsson considered,
\b \label{fun13}
f_{k}=\frac{A_{k}}{\sqrt{2k}}(1-\frac{i}{k\eta})e^{-ik\eta}+\frac{B_{k}}{\sqrt{2k}}(1+\frac{i}{k\eta})e^{ik\eta},
\e \\
and
\b \label{fun14}
g_{k}=\sqrt{\frac{k}{2}}{A_{k}}e^{-ik\eta}-\sqrt{\frac{k}{2}}{B_{k}}e^{ik\eta},
\e and used the following condition,
 \b
\label{cof115} |A_{k}|^{2}-|B_{k}|^{2}=1,
\e
after some straightforward calculation, he obtained the
coefficients as follows:
 \b \label{cof16}
 B_{k}=\alpha_{k}A_{k}e^{-2ik\eta_{0}},\;\; |A_{k}|^{2}=\frac{1}{1-|\alpha_{k}|^{2}},\e
 where
  \b \label{alf17}
  \alpha_{k}=\frac{i}{2k\eta_{0}+i},
  \e

where $ \eta_{0}=-(\Lambda/{Hk}) $ has a finite value. According to
\cite{Dan6}, we can define a finite $ \eta_{0} $
in a way that the physical momentum corresponding to $ k $ is given
by some fixed Planck scale say as $ \Lambda $. Similarly, for
this general mode function obtained:
\b
\label{two18}\langle{\phi^{2}}\rangle=\frac{1}{(2\pi)^{3}}\int
d^{3}{k}\bigg[(\frac{1}{2ka^{2}}+\frac{H^{2}}{2k^{3}})-\frac{H^{3}}{2\Lambda
k^{3}}\sin(\frac{2\Lambda}{H})\bigg],
\e
 where $
({\Lambda}/{H})\gg1 $ has been assumed. Again the first term
coming from the vacuum fluctuations in Minkowski space-time that
can be eliminated after the renormalization procedure. For example $ k=ap $, and $
p=\Lambda $, after doing some easy calculations, the power spectrum
for scalar field fluctuations in this case is obtained as
\cite{Dan6, Soj24}
\b \label{pow19}
P_{\phi}=(\frac{H}{2\pi})^{2}\Big(1-\frac{H}{\Lambda}\sin(\frac{2\Lambda}{H})\Big).
\e which is a scale-dependent power spectrum and corrections are of
order ${H}/{\Lambda}$. We will generalize the above method to our proposed excited mode.

\section{Departure from Bunch-Davies mode to the excited-de Sitter mode}
In this section, we will introduce the excited-dS mode as a fundamental mode function during inflation. But the
important question is: why excited-dS mode? Because we do not know anything about the physics of the before inflation at very early universe, but we know that cosmic inflation can be described by nearly dS space-time. So, any primary excited dS mode can be considered as a good and acceptable mode for initial state. Of course, this can be a first step towards to create of the non-linear sources of primordial fluctuations for generation anisotropy in CMBR.\\
    For the QFT in flat space-time, the vacuum expectation value of the
energy-momentum tensor becomes infinite which is removed by the
normal ordering. However in curved space-time, following trick is
used (equ. (4.5) in \cite{Bir16}),
\b \label{vac20}
\langle\Omega|:T_{\mu\nu}:|\Omega\rangle=\langle\Omega|T_{\mu\nu}|\Omega\rangle-\langle0|T_{\mu\nu}|0\rangle,
\e where $ |\Omega\rangle $ is the vacuum state in curved space-time
whereas $ |0\rangle $ stands for the  vacuum state of Minkowskian
flat background. One can interpreted the mines sign at the above
equation as the effect of the background solutions. Note that the symmetry of the curved
space-time vividly breaks in this renormalization scheme, because the background solution is not the solutions of the wave equation in
the curved space-time. Indeed, in this renormalization procedure the vacuum is defined in global curved space-time while the singularities are removed
in local flat space-time. But, if such divergences are removed by the quantities which are
defined in the curved space-time, the symmetry would be returned to theory. In this case, the background solutions are the solutions
of the wave equation in the curved space-time \cite{man17, Moh23}. By consideration this new scheme of renormalization; theory preserve symmetry of curved space-time. In addition, By having this new mode which certainly can not be dS mode, we can obtain energy-momentum tensor without any additional handmade cutoff.\\
Now let us, in the first approximation, we consider exact dS mode function as the background(BG) solution and the excited- dS mode as the fundamental solutions of the curved space-time. This excited-dS solution must be asymptotically approaches to dS background. Such an approximate mode might be obtained by expanding the Hankel function in (\ref{gen5}) up to its third term \cite{vac19, man17, Abr26} and then one can write:
\b \label{mod21}
u_{k}^{exc}\simeq\frac{1}{\sqrt{2k}}\Big(1-\frac{i}{k\eta}-\frac{1}{2}(\frac{i}{k\eta})^{2}\Big)e^{-ik\eta}.
\e
Noting that according to the proposed ansatz, the third term in mode (\ref{mod21}), is an approximate term that is added due to the expansion of the Hankel function according to (\ref{gen6}).
In the large value of $|k\eta|$, we can consider $(\frac{1}{k\eta})^{2}\rightarrow{0}$, so as a result the exited mode(\ref{mod21}), approaches the pure dS background mode as:
\b
u_{k}^{BG}=u_{k}^{BD}=\frac{1}{\sqrt{2k}}(1-\frac{i}{k\eta})e^{-ik\eta}.\e
We call solution (\ref{mod21}) as \emph{excited- dS mode} \footnote{We call this as "excited- dS mode" because in our approach, similar to "Background Field Method", we consider dS mode as the background state field and the new mode as the excited state fields in the curved space-time.}. In \cite{man17}, for the first time, we used this excited solution with the auxiliary fields to calculate the finite and renormalized power spectrum of primordial fluctuations. Also, in the our recent work \cite{man18}, we considered the Planck results (2013) for scalar spectral index of inflation \cite{obs28}, and we showed that the $\mu$ must be greater than the 3/2 and this important result stimulate us departure from linear BD mode to the non-linear excited modes.\\
In addition to the above motivations, corrections obtained from previous conventional methods for power spectrum is typically of the order of 1, 2 or higher. So it is useful for us to extend the mode up to non-linear order of its parameters. This non-linearity of our new vacuum mode appears in the conformal time variable $\eta$ . In this paper, we pursue this topic with Danielsson approach \cite{Dan6} and in the \cite{man17, man18} we have investigated it utilizes other conventional methods. It is also expected that the \emph{primordial non-Gaussianity of the CMB} come from various non-linear sources during the cosmic evolution. Non-linear term in our excited mode in the initial vacuum may leave non-Gaussian traces in the CMB \cite{Alf26, Hol29}( This issue will studied in preparing work). On the other hand, this excited mode could be more complete solution of the general wave equations (\ref{equ4}) for the general curved space-time during inflation, whereas BD mode is a specific solution for a specific dS space-time.

\section{Power Spectrum with Excited-dS Mode and Excited-$\alpha$-Vacuum}
\subsection{Calculation with excited-dS vacuum}
By inserting mode function (\ref{mod21}) in (\ref{two10}) and doing some
straightforward calculations, one obtains: $$
\langle{\phi^{2}}\rangle=\frac{1}{(2\pi)^{3}}\int
d^{3}{k}[\frac{1}{2ka^{2}}+\frac{H^{2}}{k^{3}}+\frac{a^{2}H^{4}}{8k^{5}}]
$$ The first term is the usual contribution from vacuum fluctuations
in ds space-time that can be eliminated after the renormalization.
Then  \b
\langle{\phi^{2}}\rangle=\frac{1}{2\pi^{2}}\int\frac{dk}{k}(H^{2}+\frac{a^{2}H^{4}}{8k^{2}}).
\e The power spectrum is given by \b
P_{\phi}(k)=(\frac{H}{2\pi})^{2}\Big(2+(\frac{H}{2\Lambda})^{2}\Big),
\e which is scale-dependent and the correction is of order $
({H}/{\Lambda})^{2} $, where $H$ is the Hubble parameter during inflation and $\Lambda$ is the Planck energy scale. Note that in \cite{Kal11, Kem27} with
effective field theory approach, similar order of correction has
been obtained.
\subsection{Calculation with Excited-$ \alpha$-vacuum}
 Inflation starts in approximate-dS space-time. Basically in this high energy area of very early universe with varying $H$, finding a proper mode is difficult. We offer the excited-dS solution (\ref{mod21}) as the fundamental mode during inflation that asymptotically approaches to dS background. Actually, the excited-dS mode (\ref{mod21}) can be considered as a nearest solution to dS mode. So according to (\ref{fun13}), the general solution of the equation of motion in this approximate-dS space-time, include positive and negative frequency can be given by, $$
\upsilon_{k}=\frac{M_{k}}{\sqrt{2k}}\Big(1-\frac{i}{k\eta}-\frac{1}{2}(\frac{i}{k\eta})^{2}\Big)e^{-ik\eta}$$
\b
+\frac{N_{k}}{\sqrt{2k}}
\Big(1+\frac{i}{k\eta}-\frac{1}{2}(\frac{-i}{k\eta})^{2}\Big)e^{ik\eta},
\e and call it as \emph{excited-$\alpha$- vacuum}. Note that, the mode functions (\ref{fun13}) is the general exact solution of equation(4) for pure dS space-time, and similarly we consider mode (25) as the general approximate solution of equation(4) for approximate-dS space-time. Since this vacuum
is of order $(\frac{1}{k\eta})^{2}$ , for $g_{k}$ we offer two
different choices of orders $(\frac{1}{k\eta})^{1}$ and
$(\frac{1}{k\eta})^{0}$.\\
\subsubsection{First choice of $g_{k}$}
Similar to Danielsson work, If we consider
the $g^{(1)}_{k}$, corresponding to the first derivative or
conjugate momentum of $\upsilon_{k}$ we will have, \b \label{g1}
g^{(1)}_{k}=\sqrt\frac{k}{2}{M_{k}}(1-\frac{i}{k\eta})e^{-ik\eta}-\sqrt\frac{k}{2}{N_{k}}(1+\frac{i}{k\eta})e^{ik\eta},
\e We follow the section (2.3) in \cite{Dan6} and we obtain for (25)
and (26), \b \label{cofi11}
N_{k}=-\gamma_{k}M_{k}e^{-2ik\eta_{0}},\;\;
|M_{k}|^{2}=\frac{1}{1-|\gamma_{k}|^{2}},\e where \b
\gamma_{k}=\frac{1}{4(k\eta_{0})^{2}+4ik\eta_{0}+1},\e If we ignore
the terms of higher than second order, we obtain for $\eta_{0}=-(\Lambda/Hk)$, the corrected power
spectrum as, \b
P^{(1)}_{\phi}(k)=(\frac{H}{2\pi})^{2}\Big(1-\frac{i}{2}(\frac{H}{\Lambda})^{2}\cos(\frac{2\Lambda}{H})\Big),
\e Here, the correction in (29) is of order
$(H/{\Lambda})^{2}$. \\
\subsubsection{Second choice of $g_{k}$}
In the other hand, conventionally, the vacuum is chosen by requiring that the mode
functions $\upsilon_{k}$ reduce to the Minkowski ones in the limit
$\eta\rightarrow-\infty$. So, similar to Danielsson work \cite{Dan6}, and as the
second choice, we consider $g^{(2)}_{k}$ in flat space-time as follows,

 \b \label{gen30}
g^{(2)}_{k}=\sqrt\frac{k}{2}{A_{k}}e^{-ik\eta}-\sqrt\frac{k}{2}{B_{k}}e^{ik\eta},
\e

and for (25) and (30) we again obtain $N_{k}$ and $M_{k}$
 similar to (27) but for $\gamma_{k}$ we have,
 \b \gamma_{k}=\frac{2i(k\eta_{0})+1}{4(k\eta_{0})^{2}+2i(k\eta_{0})+1},\e
and the power spectrum is given as follows,
 \b
P^{(2)}_{\phi}(k)=(\frac{H}{2\pi})^{2}(\frac{1}{1-(\frac{H}{2\Lambda})^{2}})\Big(1+
(\frac{H}{2\Lambda})^{2}-\frac{H}{\Lambda}\sin(\frac{2\Lambda}{H})\Big),
\e
If we use the following Taylor expansion for $x=\frac{H^{2}}{4\Lambda^{2}}\ll1$,

\b \sum x^{n}=\frac{1}{1-x}, \e
and if ignore the terms of higher than second order, we obtain
corrected power spectrum as,
$$P^{(2)}_{\phi}(k)=(\frac{H}{2\pi})^{2}\Big(1-\frac{H}{\Lambda}\sin(\frac{2\Lambda}{H})+
\frac{1}{2}(\frac{H}{\Lambda})^{2}$$
\b
-\frac{1}{4}(\frac{H}{\Lambda})^{3}\sin(\frac{2\Lambda}{H})+...\Big),
\e  This final result (34), includes linear and non-linear order
of trans-Planckian corrections $ {H}/{\Lambda}$. Martin and Brandenberger
in equation (79) of paper \cite{Mar14} are obtained a similar set
of corrections (29) and (34) up to second order of $ {H}/{\Lambda}$.
\\
\section{Conclusions}
In this paper, we have calculated higher order trans-Planckian corrections of power spectrum with excited-dS solution as the fundamental mode
function during inflation. This consists essentially of expanding the Hankel function for the quantum mode in dS space to
quadratic order in $\frac{1}{k\eta}$ before quantization, which corresponds to performing the quantization at finite wavelength,
rather than fully in the ultraviolet (i.e. Bunch Davies) limit. This non-trivial initial mode to be more logical since the curved space-time
symmetry has been preserved after renormalization procedure.\\
 In this excited- dS vacuum, slightly deviation of the exact solution lead to a correction to the power spectrum which
is of non-linear order $ {H}/{\Lambda}$, where $H$ is the Hubble parameter and  $ \Lambda $ is the fundamental energy scale of the theory during inflation in very early universe. The corrections to the power spectrum that obtained with this alternative mode is
more complete than the corrections obtained from conventional method with pure dS mode, that of course in the dS limit leads to standard result. Finally, due to
selection of this excited mode, one expects creation of particles and non-Gaussianity of CMBR,  which will study in the future works. \vspace*{4mm}

\begin{acknowledgments}
The authors would like to thank M.V. Takook, M. R. Tanhayi and P. Pedram for useful comments. This work has been supported by the Islamic Azad University, Science and Research Branch, Tehran, Iran.
\end{acknowledgments}

\end{document}